# Fiber-to-Waveguide and 3D Chip-to-Chip Light Coupling Based on Bent Metal-Clad Waveguides

Zhaolin Lu, Peichuan Yin, and Kaifeng Shi

*Microsystems Engineering*
*Kate Gleason College of Engineering*
*Rochester Institute of Technology, Rochester, New York, 14623, USA*
zhaolin.lu@rit.edu

**Abstract**

Efficient fiber-to-waveguide light coupling has been a key issue in integrated photonics for many years. The main challenge lies in the huge mode mismatch between an optical fiber and a single mode waveguide. Herein, we present a novel fiber-to-waveguide coupler, named "L-coupler", through which the light fed from the top of a chip can bend 90° with low reflection and is then efficiently coupled into an on-chip Si waveguide within a short propagation distance (<20$\mu$m). The key element is a bent metal-clad waveguide with a big matched input port. According to our finite-difference time-domain (FDTD) simulation, the coupling efficiency is over 80% within a broad range of working wavelengths in the near-infrared regime for a transverse electric input Gaussian wave. The coupler is polarization-dependent, with very low coupling efficiency (6%-9%) for transverse magnetic waves. The coupler can also be used for three-dimensional (3D) chip-to-chip optical interconnection by efficiently coupling light into an integrated photonic chip from a vertical-cavity surface-emitting laser (VCSEL), or Si waveguide on another chip. The work presented in this Arxiv article has been filed as a patent [1]. Detailed description of this work may be published in future papers.

## 1. Introduction

Fiber-optical links have successfully enabled tera-bits-per-second transcontinental communications. With the rapid development of nanoscale photonic and plasmonic devices, recent years have seen the advancement of integrated photonics, which may bring the advantages of photonics into the very-large-scale integrated (VLSI) circuits and realize on-chip communication and computation at a tremendous speed. The future Internet of Things (IoT) requires both long-distance optical fiber links and on-chip ultrafast optical communications. However, there has been a long-standing problem to connect an optical fiber network with an on-chip photonic integrated circuit. Simply speaking, efficient fiber-to-waveguide light coupling has not been well established.

The main challenge lies in the huge mode size mismatch between an optical fiber and a single-mode semiconductor waveguide. One main advantage of semiconductors, e.g. Si, for photonic applications is their large refractive indices at telecom wavelengths, which can confine light within nanoscale dimensions. A typical single mode Si waveguide has a height, 250±50 nm, and a width, 400±100 nm. Further increase of the dimensions will introduce additional modes and the waveguide will not work at the single mode any more. On the other hand, the core diameter of a single mode fiber is about 8$\mu$m-10$\mu$m. Thus, the overall cross-section area mismatch is over 500:1, which leads to a very poor efficiency to directly couple light from a single mode fiber into a Si waveguide, with a typical loss over 20 dB. Several techniques have been developed including grating coupling [2,3,4], prism coupling [5], and vertical tapering [6], but they are either difficult in fabrication or with poor efficiency. During recent years, some other novel couplers were also reported, including a tapered graded index structure [7], 45°-micro-mirrors [8], laser-induced fiber gratings [9], inverted taper [10,11], and evanescent couplers [12]. However, none of the approaches provides sufficient efficiency for coupling into sub-micrometer semiconductor waveguides within a compact dimension. The state-of-the-art solution remains the inverted taper, but it requires a very long coupling length to achieve high efficiency, as well as edge polishing and deep V-wedge etching for mounting. Herein, we report a fiber-to-waveguide coupler based on a bent metal-clad waveguide, which is named "L-coupler" by us for now.

## 2. L-Couplers

Basically, light is first vertically fed into a "horn waveguide" and is then funneled into a "throat waveguide". Both horn and throat waveguides are hollow waveguides with metal shells. The key element is a bent waveguide, which is composed of the throat waveguide and a metal-sandwiching $Si_3N_4$ waveguide. The bent waveguide can steer the light propagation 90°, from the vertical direction to the horizontal direction, and couple light into the metal-sandwiching $Si_3N_4$ waveguide metal-sandwiching $Si_3N_4$ waveguide on chip. Finally, the light in the $Si_3N_4$ waveguide is coupled into a single mode Si waveguide through double inverted tapers.

Bent metal-clad waveguides have been used in microwave circuits for decades [13]. They can steer the light propagation in waveguide 1 to waveguide 2 for 90°, as shown in Fig. 1(a). To decrease the reflection, two basic rules need to follow: (1) waveguide 1 should work at the single mode; (2) the parameters of these two waveguides should be optimized. In microwave circuits, metal is treated as perfectly electric conductor, whereas metal absorption needs to be considered in the near-infrared (NIR) regime. In 2D configuration, a metal-clad waveguide is a metal-insulator-metal (MIM) waveguide, which can always support at least one plasmonic transverse magnetic (TM) mode in the NIR regime regardless how small the channel is. Bent MIM plasmonic waveguides, which work at TM modes, have been extensively discussed [14,15]. To ensure low reflection, the channel width $b$ usually reduces to 50-100 nm. The nanoscale width has advantages for the integration of ultracompact waveguides. On the other hand, the non-plasmonic transverse electric (TE) mode allows a larger single mode channel. The allowed channel width $b$ for a 2D hollow single TE mode is in the range of $\frac{\lambda_o}{2} \leq b < \lambda_o$, where $\lambda_o$ is the working wavelength. Thus, the hollow waveguide can still operate at single mode even if the channel width goes up to 1 $\mu$m or more. The large waveguide width allows for a broader in-coupling beam. To this end, we focused our work on the TE mode. According to our best knowledge, this is first work proposing the applications of bent metal clad waveguides in fiber-to-waveguide and 3D chip-to-chip light coupling.

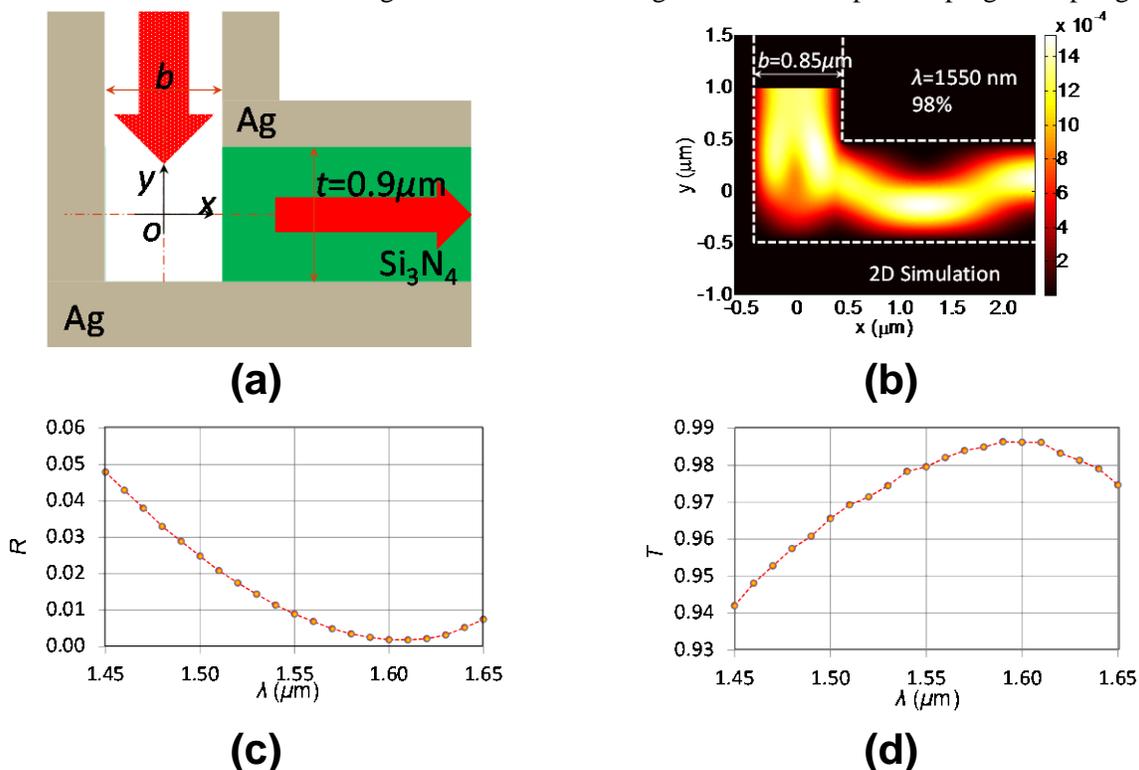

Figure 1. (a) The illustration of the bent metal-clad waveguide. (b) The FDTD simulation of the power flow. (c) The power reflectance as a function of the working wavelength. (d) The power transmittance as a function of the working wavelength.

Another novelty of the L-coupler is that the involved waveguides in the bend are filled with media with different refractive indices. To decrease the attenuation, Ag is used as the cladding metal in the coupler. In our work, waveguide 1 is a hollow waveguide; waveguide 2 is a Ag-clad $Si_3N_4$ waveguide and the thickness of $Si_3N_4$ is set as $t = 0.9$ μm as shown in Fig. 1(a). We considered the light coupled from the matched TE mode in waveguide 1 into waveguide 2. The width $b$ of the hollow waveguide is optimized by sweeping different $b$ values and find that the maximum coupling efficiency up to 98.0% can be achieved when $b = 0.85$ μm. The simulation is performed for 2D modelling. The refractive index of $Si_3N_4$ is assumed 1.98 at $\lambda_o = 1550$; the complex dielectric constant of Ag is $\epsilon_r = -129.17 + j3.28$ at $\lambda_o = 1550$ nm based on Johnson and Christy's work [16]. The value is close to those obtained in recent works [17-19]. Input is located at 1 μm from the bending point [the origin $O$ is shown in Fig. 1(a)] and output detector is placed in waveguide 2 at $x = 2$ μm, while the reflectance is only 0.9% at $\lambda_o = 1550$ nm, measured 40 nm away from the source. The 1550-nm light source is assumed as the matched mode of the hollow waveguide. Figure 1(b) shows the FDTD simulation of the power density distribution when a matched mode is fed in the hollow waveguide. As can be seen, most power is coupled from waveguide 1 into waveguide 2. In this sense, the bent waveguide can be treated as a good 45° tilted mirror. Very high transmittance can be achieved in a large wavelength range, as shown in Fig. 1 (c, d). Even higher transmittance and lower reflectance can be obtain at $\lambda_o = 1600$ nm.

A similar result can be achieved when the guiding air in waveguide 1 is replaced by $SiO_2$ with refractive index $n = 1.44$. In this case, the maximum efficiency is 94.3% when $b = 0.6$ μm. In addition, the filling material in waveguide 2 does not have to be $Si_3N_4$. Other materials will also work well, for example $Al_2O_3$, $TiO_2$, which have larger indices than the filling material in waveguide 1. The physical mechanism is still under investigation, and the results will be reported in a different work.

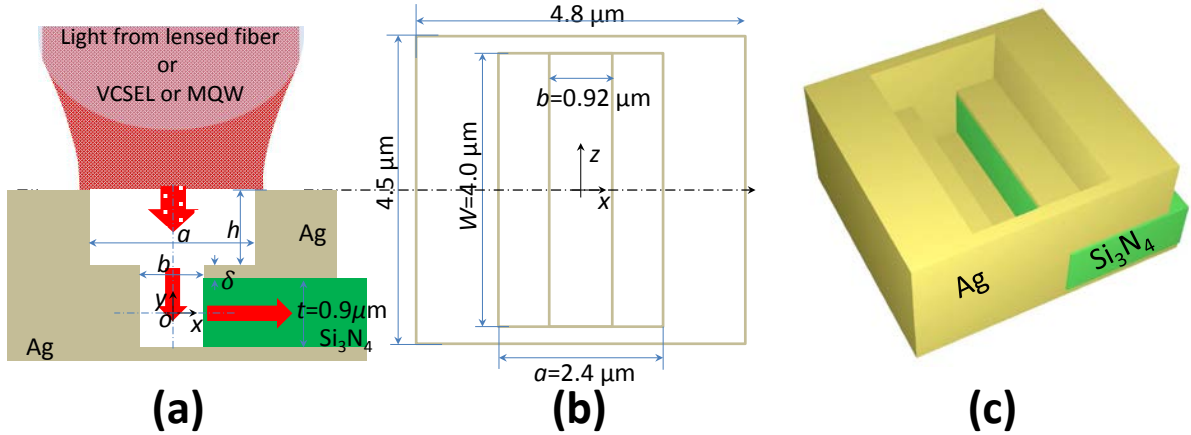

Figure 2. The illustration of the input port and bent waveguide: (a) Sectional view (cut from the center of the coupler). (b) Top view. (c) Perspective view.

To allow a broader input beam, a horn is added as the input port as shown in Fig. 2(a). If the width, $a$, and depth, $h$, of the horn are carefully designed, a broader input beam can funnel into a narrower waveguide. The mechanism is still not very clear, but the surface current might play an important role. Note that the surface current, induced by the electric field component of light for the TE mode, is along the waveguide sidewall direction. Furthermore, the 2D simulation results are also roughly valid even if the dimension in the $z$-axis is finite. Assume that the input is a Gaussian beam with field diameter $2r = 3$ μm, which may come from a lensed fiber or VCSEL. We set the width of the input horn $W = 4.0$ μm, and swept the values of $a$ and $h$ in a large range. Three dimensional FDTD simulation shows that coupling efficiency can go up to 89.3% at $\lambda_o = 1550$ nm when $a = 2.4$ μm and $h = 1.1$ μm for a Gaussian beam with diameter $2r = 3$ μm fed into the horn. Figure 2(b) marks the parameters of the L-coupler based on the sweeping optimization. Figure 2(c) illustrates the perspective view of the L-coupler. To decrease the loss by the plasmonic mode, the metal-sandwiching $Si_3N_4$ waveguide does not have side walls, but only has metal clad layers at the top and bottom. The dimension in the *z*-direction is 4.5 $\mu m$. As can be seen, the throat width is slightly adjusted to $b = 0.92\,\mu m$ for better 3D performance. The thickness of the Ag coating layer over the $Si_3N_4$ layer is $\delta = 100$ nm.

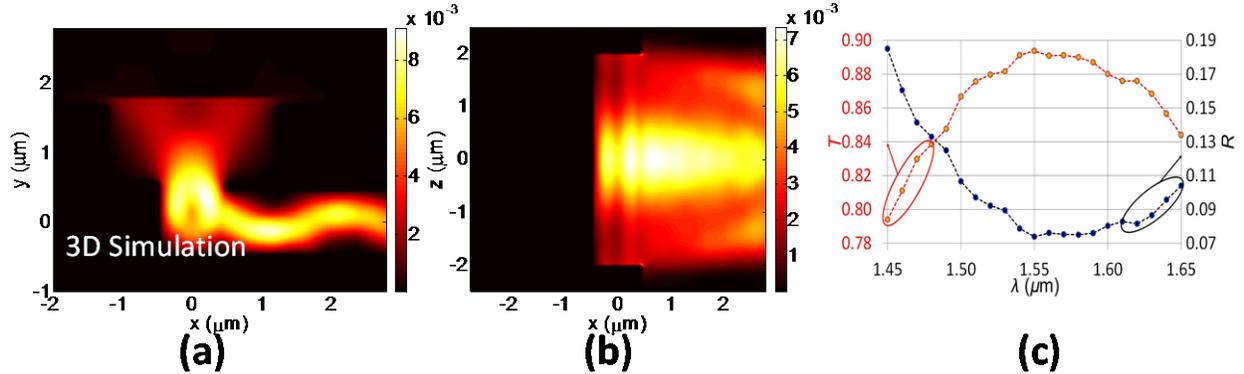

Figure 3. The 3D FDTD simulation of light-coupling into the L-coupler: (a) Power distribution in the *xoy* plane. (b) Power distribution in the *xoz* plane. (c) The power transmittance and reflectance as a function of wavelength.

Figure 3 shows the 3D simulation results of the light intensity distributions when the Gaussian beam ($2r = 3\,\mu m$) is coupled into the bent waveguide. Figure 3(a) shows the intensity distribution in the in $z = 0$ plane; Fig. 3(b) shows the intensity distribution in the $y = 0$ plane, which goes through the center of the metal-sandwiching $Si_3N_4$ waveguide. As can be seen in Fig. 3(c), over 80% power transmittance can be achieved within a large wavelength range. In particular, power transmittance up to 89.3% can be achieved at 1550 nm. When the wavelength spans away from the center wavelength, the transmittance decreases, mainly due to the increase of reflectance, since the horn waveguide functions as a very low-Q resonator. To shift the center wavelength, the dimensions of the horn waveguide should be redesigned.

## 3. Coupling into Si Waveguides

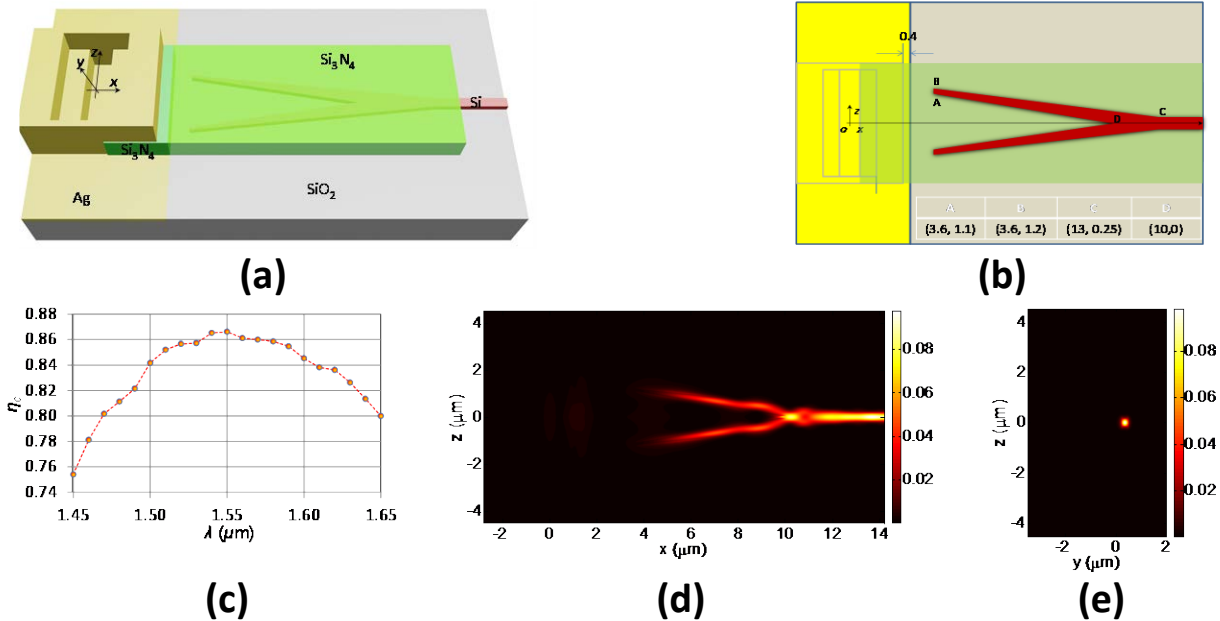

Figure 4. (a) The illustration of the "LV-coupler". (b) The location and vertices of the double inverted taper, which is symmetric to the *x*-axis. The vertex (*x, z*) coordinates (unit: $\mu m$) are listed in the insert. (c) The coupling efficiency as a function of working wavelengths. (d) The power distribution in *xoz* plane through the center of the Si waveguide. (e) The power distribution in the *yoz* plane at the output port.

To avoid more metal absorption, the metal-sandwiching waveguide is terminated and then light propagates in a $Si_3N_4$ waveguide with a cross-section 0.9μm-by-4.5μm, which also simultaneously works as the cladding layer of the Si waveguide. Importantly, the mode profiles of the metal-sandwiching $Si_3N_4$ waveguide and "nude" $Si_3N_4$ waveguide can well match for TE modes, and a very large portion of power is coupled into the latter one. A slight mode mismatch exists at the bottom $Si_3N_4$-$SiO_2$ boundary. To decrease the scattering, the bottom metal layer is designed 0.4 μm longer than the top layer.

To collect light in a large range in the transverse direction, two inverted tapers are arranged in the form of a V-shape is employed as shown in Fig. 4 (a, b). The thickness and width of the Si waveguide are assumed to be 240 nm and 500 nm, respectively. In Fig. 4(a), the $Si_3N_4$ over the V-coupler is same as that in the metal-clad waveguide. The transparent color is used to easily illustrate the V-coupler below the clad. The coupling efficiency is up to 86.4% based on this "LV-configuration" at $\lambda_o = 1550$ nm. The coupler can work in a broad wavelength range. Figure 4(c) plots the coupling efficiency as a function of working wavelength. As can be seen, over 80% coupling efficiency can be achieved within the working wavelength range of 1.47-1.64 μm. Note that there is still about 3% loss between the power in the Si waveguide and that from the L-coupler. In our configuration, the loss comes from three parts: bottom scattering, 1.8%; top scattering, 0.5%; side scattering, 0.7%. Various approaches are available to couple light in the $Si_3N_4$ coating into a single mode Si waveguide. Some of them may further improve the coupling efficiency.

Figures 4(d&e) show the simulation of the power distributions at $\lambda_o = 1550\ nm$ in different planes. According to the power distribution in the *yoz* plane [Fig. 4(c)] at the output, the double V-tapers can efficiently collect the power in the $Si_3N_4$ cladding into the nanoscale Si waveguide. In the cladding layer, there is almost no power outside of the Si waveguide. The coupler is very sensitive to the polarization. When the TE-polarized Gaussian beam is switched to TM-polarized one as the input, the coupling efficient sharply decreases to 6%-9%.

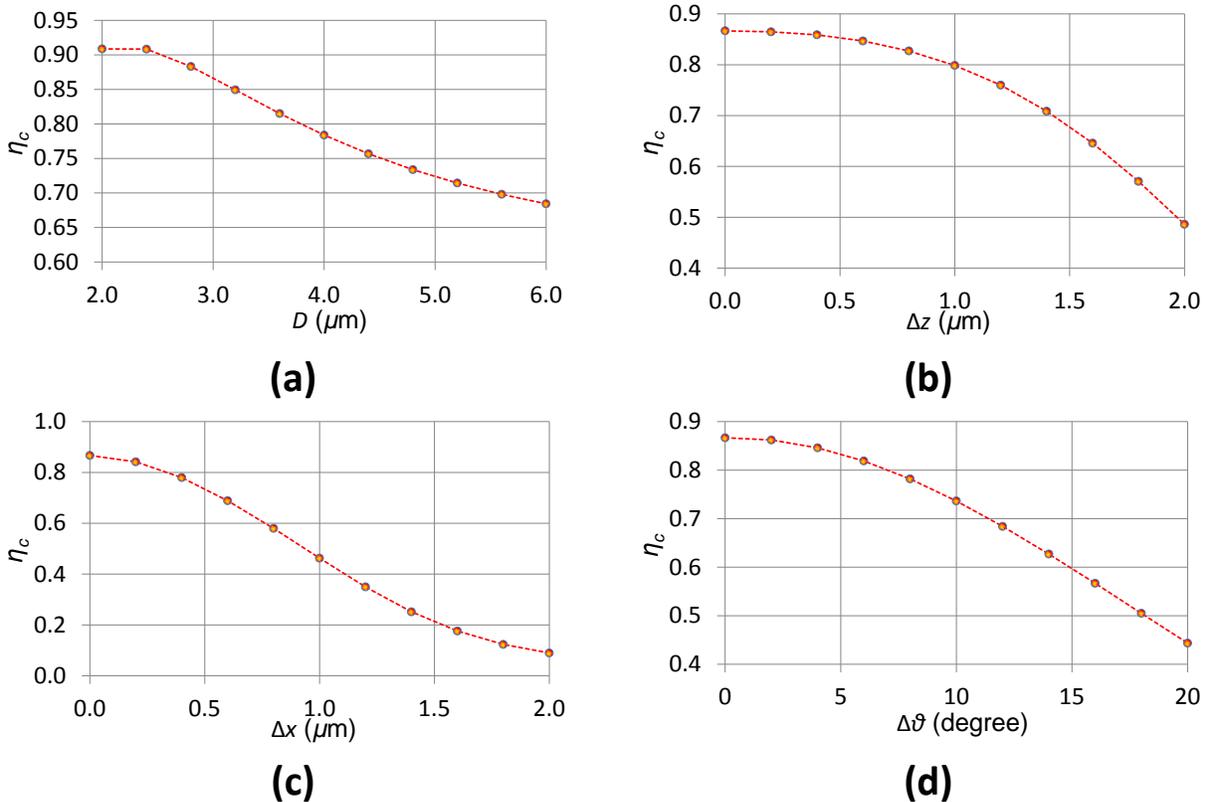

Figure 5. The alignment tolerance of the coupler: (a) Beam diameter tolerance. (b) Transverse shift in *z*-direction. (c) Transverse shift in *x*-direction. (d) Angle tolerance.

The L-coupler allows for considerable alignment tolerance. Figure 5 shows the coupling efficiency as a function of beam size, transverse alignment, and angle tolerance. As shown in Fig. 5(a), the smaller the incident beam size, the greater the coupling efficiency. The coupling efficiency can increase up to 91% if the beam size is squeezed into $2r = 2.4$ μm or less, which is actually not easy to achieve. Thus, we assume that the beam size is $2r = 3.0$ μm throughout this work. On the other hand, the coupling efficiency is not sensitive to the beam size. When the beam size increases to $2r = 4.0$ μm, and $2r = 6.0$ μm, the corresponding coupling efficiencies are 78% and 68%. As shown in Fig. 5(b, c), the 3dB-tolerances in the shift in *z*- and *x*- directions are ±2 μm and ±1 μm, respectively, which are similar as that of the inverted taper. In addition, the L-coupler is not sensitive to the incident beam direction. As shown in Fig. 5(d), the 3dB-tolorance in the angle is over ± 15°.

## 4. Three-Dimensional Chip-to-Chip Coupling

Three-dimensional integrated circuits have been a key approach to improve the performance of next generation VLSI. Currently available techniques for 3D chip-to-chip optical interconnection are free-space or waveguide links based on mirrors, prisms, or gratings [20- 25], which are difficult for fabrication and have large insertion loss. In addition to couple light from an optical fiber or VCSEL, the L-coupler can also be used to realize 3D chip-to-chip interconnection where two chips placed face-to-face are optically linked. In this case, L-coupler functions as a 45° tilted mirror or prism. As examples, Fig. 6 shows the illustration and simulation. As can be seen, light from a Si waveguide in one chip can be coupled vertically into the Si waveguide in another chip. The propagation directions of the Si waveguides in different chips can be either in the same direction or opposite directions. Our preliminary simulation shows that coupling efficiency between 75%-80% may be achieved. The detailed description on the parameters will be discussed in future publications.

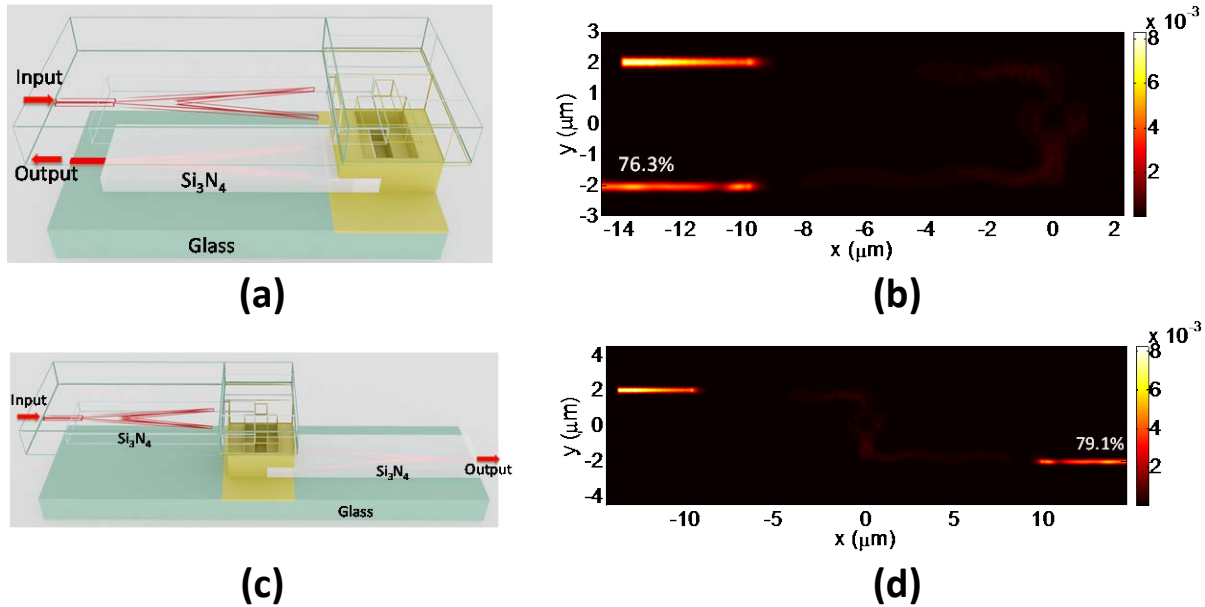

Figure 6. Application of L-couplers in 3D chip-to-chip coupling: (a, b) The layout and simulation 3D chip-to-chip optical coupling based on two LV couplers, where the light input and output propagate in opposite directions. (c, d) The layout and simulation 3D chip-to-chip optical coupling based on two LV couplers, where the light input and output propagate in the same direction.

When amorphous Si is used, the Si tapers can be inserted into metal-clad glass waveguides as shown in Fig. 7. Coupling efficiency may be improved to about 85% [Figs. 7(a-d)]. In addition, very efficient power splitting can be simultaneously achieved during the 3D chip-to-chip optical coupling. See Fig. 7 (e, f). If the distance between the two chips is over 10 μm, a dielectric waveguide or optical fiber can be used to link the two L-couplers, as shown in Fig. 7(g, h). Also, the detailed description on the parameters will be discussed in future publications.

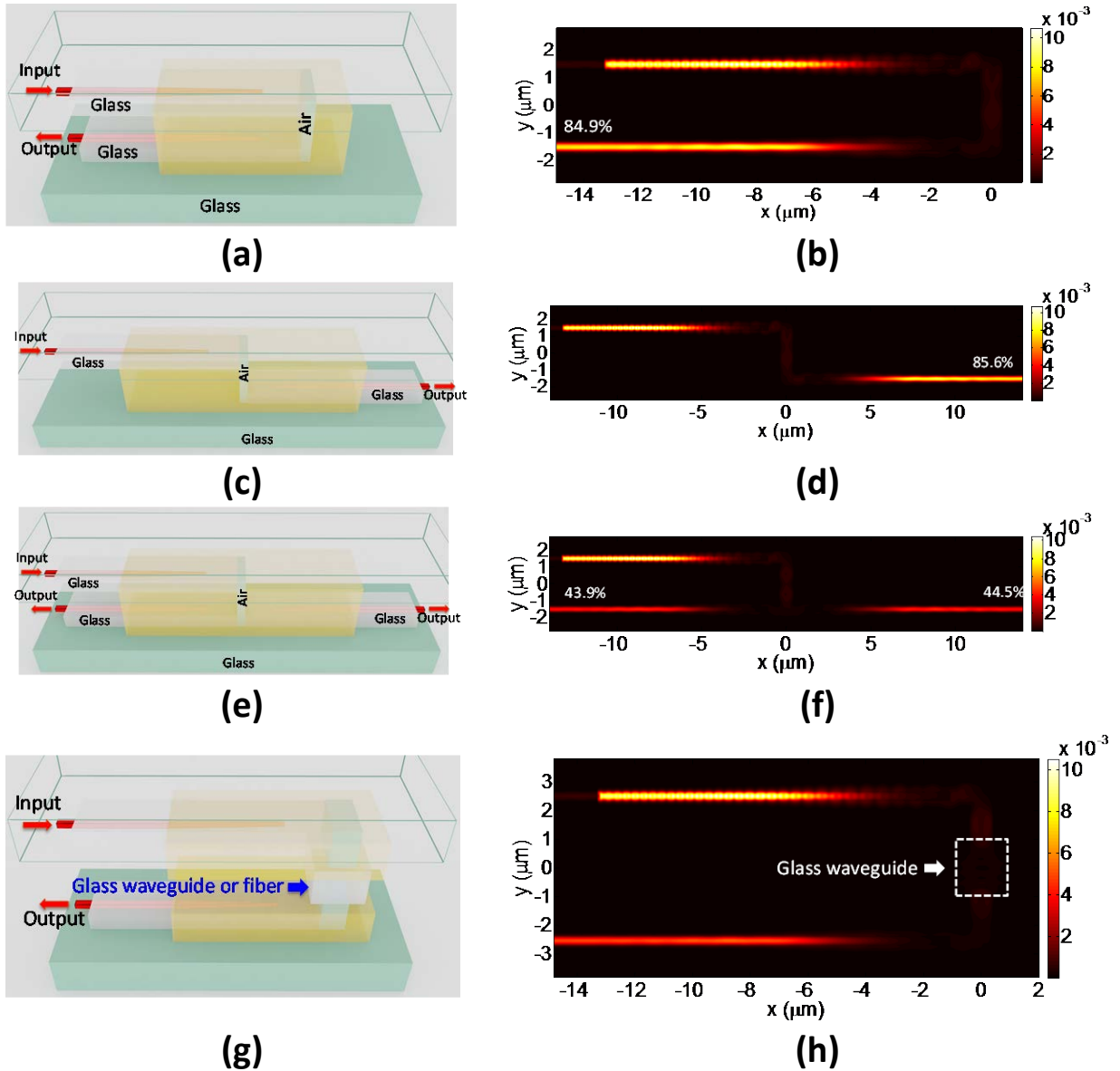

Figure 7. Application of L-couplers in 3D chip-to-chip coupling, where Si tapers are inserted in metal-clad waveguides: (a-d) The illustration and simulation of L-couplers for 3D chip-to-chip coupling, where metal-clad waveguides are used to link the two L-couplers. (e, f) The illustration and simulation of L-couplers simultaneously for 3D chip-to-chip coupling and power splitting. (g, h) The illustration and simulation of L-couplers simultaneously for 3D chip-to-chip coupling, where a dielectric waveguide or optical fiber is used to link the two L-couplers for long chip distance.

## 5. Discussion

The LV-coupler as shown in Fig. 4(a) can be fabricated in the following steps: (1) SOI waveguide fabrication. (2) Etch 100-nm oxide and deposit 100-nm Ag. (3) Deposit 0.9 $\mu m$ $Si_3N_4$. (4) Deposit 0.9 $\mu m$ Ag. (5) Deposit 100 nm Ag. (6) Deposit 1.1 $\mu m$ Ag. These fabrication processes may result in the device as shown in Fig. 8(a), which is slightly different from the one shown in Fig. 4(a). Nevertheless, the effect of the extra $Si_3N_4$ structure on coupling efficiency is negligible (the Si waveguide is assumed 240 nm thick) according to the simulation. Furthermore, one Ag layer may be saved with the tradeoff ~0.5% coupling efficiency decrease at 1550 nm. See Fig. 8(b).

The cladding Ag may be replaced by Cu, which is CMOS compatible. According to two recent works [17, 18], the dielectric constant of Cu at 1550 nm $\epsilon_r = -119.23 + j4.2446$, and $\epsilon_r = -119.87 + j4.080$, respectively, is quite close to that of Ag. If these dielectric constants of Cu are used in the simulation, the coupling efficiency will decrease about 0.5%.

One main drawback of the L-couplers is that they work well only for TE (nonplasmonic) modes. For TM (plasmonic) modes, the coupling efficiency is only 6%-9%. However, this polarization-dependent feature is not always a drawback. For one thing, almost no integrated photonic circuit is required to simultaneously work for both TE and TM modes, and many devices can be designed for TE modes. In contrast, the drawback may become an advantage for TE modes because the couplers can be treated as devices naturally integrated with polarizers. Furthermore, once a TE mode is coupled in an on-chip waveguide, it can be converted into a TM mode based on some waveguide devices.

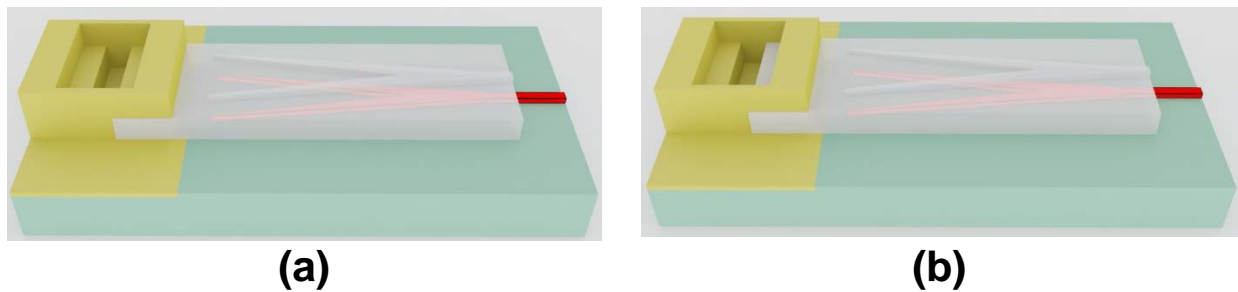

**(a)**  **(b)**

Figure 8. (a) The resulting device during fabrication. (b) A varication of the proposed device for easier fabrication.

## 6. Summary

In summary, we proposed and did numerical study, for the first time, on the bent metal-clad waveguides for fiber-to-waveguide and 3D chip-to-chip light coupling applications, i.e. L-couplers. The L-couplers, each of which is made of two waveguides filled with different materials, work well for TE modes at NIR wavelengths. For the applications of the L-couplers in fiber-to-waveguide coupling, the L-couplers have the following advantages:

(1) High efficiency and broad wavelength range. Our preliminary FDTD simulation shows over 80% within $\lambda$=1.47-1.64 $\mu$m.
(2) CMOS-compatible (when Ag is replaced by Cu). The involved fabrication processes are lithography, thin film deposition, and liftoff. No edge polishing is needed because light is coupled vertically over a chip.
(3) Compact. The coupler length only takes up 20 $\mu$m or less.
(4) Allowing for quite large alignment tolerance because each L-coupler may have a big horn as its input port.

For the applications of the L-couplers in 3D chip-to-chip coupling, this technique has the following advantages in addition to high efficiency, broad wavelength range, CMOS-compatible fabrication, and compact dimensions:

(1) L-couplers function as a 45° tilted mirrors or prisms.
(2) Glass waveguide or fiber can be added as through-Si-via (TSV) for longer distance connection.
(3) Power splitting and interconnection can be realized simultaneously.

**Acknowledgments**

We would like to acknowledge the helpful discussions with Dr. Michael Gerhold from U.S. Army Research Office (ARO) and Dr. Justin Bickford from U.S. Army Research Laboratory (ARL).